\documentstyle[prb,aps,twocolumn]{revtex}
\input epsf
\epsfverbosetrue
\begin{document}
\draft
\flushbottom
\twocolumn[
\hsize\textwidth\columnwidth\hsize\csname @twocolumnfalse\endcsname

%\preprint{UBCTP-95-~~~ cond-mat/95~~~ }
\title{
Quantum Oscillations of Electrons and of Composite Fermions in Two
Dimensions: Beyond the Luttinger Expansion
} 

%%%%
%\vskip .3in
%%%%

\author{S. Curnoe$^{1,3}$ and P. C. E. Stamp$^{1,2,4}$}

%%%%
%\vskip .1in
%%%%

\address{$^1$Department of Physics \& Astronomy and $^2$Canadian Institute
for Advanced Research\\University of British Columbia,
Vancouver, B.C. V6T 1Z1, Canada\\
$^3$Department of Condensed Matter Physics, Weizmann Institute of Science,
Rehovot 76100, Israel$^{*}$\\
$^4$Institut Laue-Langevin, BP 166X, Grenoble 38042, France$^*$}
\maketitle
%%%%
%\vskip .3in
%%%%

\begin{abstract}
Quantum oscillation phenomena, in conventional  
2-dimensional electron systems and in the 
fractional quantum Hall effect, are usually treated
in the Lifshitz-Kosevich formalism.  This is justified in three 
dimensions by Luttinger's expansion, in the parameter
$\omega_c/\mu$.  We show that in two dimensions this expansion breaks down, 
and derive a new expression, exact in the limit where
rainbow graphs dominate the self-energy.
Application of our results to the fractional quantum Hall effect 
near half-filling shows very strong deviations from Lifshitz-Kosevich 
behaviour. We expect that such deviations will be important in any 
strongly-interacting 2-dimensional electronic system.

\end{abstract}


%%%%
\vskip 1cm
%%%%
]
\narrowtext
\tightenlines

Quantum oscillation (QO) phenomena (in which Landau quantisation 
causes all thermodynamic and transport properties of conductors to
oscillate with $1/B$, where $B$ is the sample induction) have been for
four decades amongst the most powerful tools in solid state
physics \cite{shoen1,ashcr1} in two and three dimensions.
The recent composite fermion (CF) theory \cite{jain1,hlr} of the
fractional quantum Hall effect (FQHE) predicts similar oscillations in $1/b$,
where $b = B- B_{1/2}$ and $B_{1/2} = 2 n_e/e$ is the mean
``statistical field'' coming from double fluxons attached to the
CF's.  Intense experimental interest in FQHE systems near half-filling 
\cite{willett} has given strong evidence for the CF theory (eg., from
Shubnikov-de Haas QO experiments \cite{du1,leadl1}, and analogous
oscillations in acoustic absorption \cite{wille1} and
compressibility \cite{eisen1}).  The oscillations have been fit using 
Lifshitz-Kosevich (LK) formulae \cite{shoen1,lifsh1}, usually with an
impurity scattering Dingle temperature (sometimes assumed energy dependent),
and an ``effective mass'' $m^{*}$. The CF cyclotron
frequency $\omega_{CF}^{*} = eb/m^{*}$ increases
rapidly with $b$ near $b = 0$ (ie., near half-filling),
because of strong infra-red divergent gauge interactions
\cite{stern1,kim1,curnoe1,curnoe2}.

The LK formulae (and generalisations of them, incorporating low energy
fluctuations \cite{fowle1,engel2}) rely fundamentally on an 
expansion of the {\it oscillatory part} of the 
thermodynamic potential $\Omega(B)$, in powers of $\omega_c/\mu$ 
(where $\omega_c = eB/m$, and $\mu$ is an upper cut-off, equal to the chemical
potential in the simplest models), given by Luttinger \cite{lutti2}.   He 
wrote the 1-particle self-energy \cite{gauge} as 
$\Sigma(B) = \bar{\Sigma} +\Sigma_{osc}(B)$,
where $\Sigma_{osc}$ contained all contributions oscillating in
$1/B$ (and analogously the fermionic Green's function
 ${\cal G} = \bar{{\cal G}} +{\cal G}_{osc}$).  Expanding the 
functional $\Omega(\Sigma)$  around $\Omega(\Sigma=\bar{\Sigma})$ in 
powers of $\Sigma_{osc}$, one finds \cite{lutti2}
\begin{equation}
\Omega = \frac{-1}{\beta}\mbox{Tr}[\log\bar{\cal{G}}^{-1} - \bar{\Sigma}
\bar{\cal{G}}]  + \Phi(\bar{\Sigma}) + O(\Sigma^2_{osc})
\label{eq:luttiexp}
\end{equation}
where $\beta = 1/k_B T$.  In three dimensions,
\begin{eqnarray}
\frac{1}{\beta}\mbox{Tr}[\log(\bar{\cal{G}}^{-1})] & \sim  &
      \left(\frac{\omega_c}{\mu}\right)^{5/2}  \label{eq:3dG} \\
\frac{\Sigma_{osc}}{\bar{\Sigma}} & \sim &\left(\frac{\omega_c}{\mu}\right)^{3/2}
\label{eq:3dsigma}
\end{eqnarray}
and also $\Phi(\bar{\Sigma}) + \beta^{-1} \mbox{Tr}(\bar{\Sigma} \bar{\cal{G}})
= 0$ at
least to $\sim O((\omega_c/\mu)^{3})$.  Thus writing $\Omega = \Omega_0 + 
\Omega_{osc}$, we have that the leading
oscillatory contribution up to $O((\omega_c/\mu)^{5/2})$ is contained in
\begin{eqnarray}
\Omega   &\sim &  \frac{-1}{\beta}\mbox{Tr}[\log \bar{\cal{G}}^{-1}] \\
 & = & \frac{-1}{\beta}
\sum_{i\omega_m,n,\sigma,k_z}    
\log\left[i\omega_m-
\epsilon_{\sigma n k_z} 
+\bar{\Sigma}(i\omega_m,\epsilon_{\sigma n k_z}) \right] 
\label{eq:3dosc}
\end{eqnarray}
with
\begin{equation}
\epsilon_{\sigma n k_z} = \epsilon_{\sigma} + 
\left(n+\frac{1}{2}\right)\hbar \omega_c +\frac{\hbar^2k_z^2}{2m} -\mu
\end{equation}
where $\mu$  is defined as the zero of the energy, $n$  labels the
Landau levels and $\sigma$ is a spin index.  
Eq. (\ref{eq:3dosc}), which contains the {\em non-oscillatory}
self-energy $\bar{\Sigma}$, provides the fundamental justification for
extracting the {\it zero field}, many-body interaction-renormalised band
structure from QO experiments \cite{shoen1}.

In this paper we show that

{\bf (a)} Luttinger's expansion {\em fails} in any interacting
2D electronic system; however

{\bf (b)} an alternative expansion can be found under certain
circumstances (see below), in which now the full self-energy
(including the highly singular $\Sigma_{osc}$) must be used.

{\bf (c)} This new expansion can give results sharply different from the 
previous ones \cite{shoen1,lifsh1,fowle1,engel2,lutti2}.

To show the practical importance of these results, we will
apply them to CF's; however they are
relevant in principle to any 2D electronic system \cite{Balthes}.

{\bf (i) Failure of Luttinger's Expansion:} We first repeat the 
analysis which  yields Eqs. (\ref{eq:3dG},\ref{eq:3dsigma}), but now
in two dimensions.  We shall find quite generally that
\begin{eqnarray}
\frac{1}{\beta} \mbox{Tr}[\log \bar{\cal{G}}^{-1}] & \sim & 
\left(\frac{\omega_c}{\mu}\right)^2 \label{eq:2dG} \\
\frac{\Sigma_{osc}}{\bar{\Sigma}} & \sim & \frac{\omega_c}{\mu}
\label{eq:2dsigma}
\end{eqnarray}
Thus the term $\sim O(\Sigma_{osc}^2)$ is 
as important as the ``leading'' term, and the whole
expansion must be re-examined.

\begin{figure}[ht]
\epsfysize=2.8in
\epsfbox[30 270 575 750]{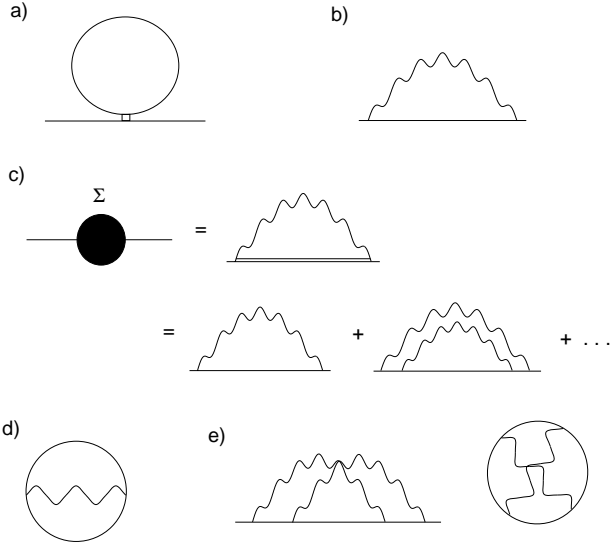}
\caption{The Feynman graphs discussed in the text. The self-energy graphs
include (a) the Hartree-Fock term, (b) the lowest order ``RPA'' term, (c) a
self-consistent ``rainbow graph'' sum of such terms. In (d) we have the lowest
order (2nd-order) contribution to $\Omega$, and (e) shows the lowest
(4th-order) ``crossed graph'' contributions to $\Sigma$ and $\Omega$.
\label{fi:feyn}}
\end{figure}
Eq. (\ref{eq:2dG}) is easily verifed.  To justify (\ref{eq:2dsigma})
we first repeat, in 2 dimensions, Luttinger's 3-dimensional
calculation of the graph in Fig.\ \ref{fi:feyn} (a); 
we then
extend the argument to higher graphs.   The graph in Fig.\ \ref{fi:feyn} (a)
has the real-space form 
\begin{equation}
\Sigma_R(\vec{r}) = \int d^3r^{'} V(\vec{r}-r^{'})
[g(0) - \hat{P}_{r r^{'}}g(r-r^{'}) e^{i\phi(r,r^{'})}]
\end{equation}
where $\hat{P}_{r r^{'}}$ is the exchange operator, $\phi(r,r^{'})$ is a
gauge-dependent phase factor \cite{lutti2}, and
$g(r) = \bar{g}(r) + g_{osc}(r)$, where
\begin{eqnarray}
\bar{g}(r) & = &\oint_{\bar{c}} dt M(r,t)  \label{eq:gbar}\\
g_{osc}(r)& =& \sum_{l} \oint_{C_l} dt M(r,t) \label{eq:gosc} \\
M(r,t) & = & \frac{-ie^{\beta\mu t}}{2 \sin \pi t} F_{2D}(r,t)   \\
F_{2D}(r,t) & = & \frac{\omega_c}{4 \pi \sinh(\beta \omega_c t/2)} \nonumber \\
& & \hspace{.5in}\times\exp\left[
\frac{-\omega_c}{4}\coth\left(\frac{\beta \omega_c t}{2}
       r^2\right)\right] \label{eq:f2}
\end{eqnarray}
The contour $\bar{C}$ encircles the negative real axis counterclockwise,
the contours $C_l$ likewise encircle the points
$T_l = 2 \pi i l/\beta \omega_c$, with
$l = \pm1, \pm 2, \ldots$.  The 3D function $F_{3D}(r,t)$ differs
from (\ref{eq:f2}) by the factor 
$(2\pi\beta t)^{-1/2}\exp[-z^2/2\beta t]$,
where $z$ is the third dimension, perpendicular to $r$ (cf.
Ref.\ [16] Eq. (A.16)). It is this difference which yields
(\ref{eq:2dsigma}), instead of (\ref{eq:3dsigma}), upon integrating over 
$t$ in (\ref{eq:gbar}) and (\ref{eq:gosc}).

Consider now graph \ref{fi:feyn}(b), assuming that the internal boson line
represents either (i) a phonon, or a conventional ``Fermi liquid''
electronic fluctuation; or (ii) a singular gauge fluctuation
\cite{stern1,kim1,curnoe1,curnoe2}.  
 Using the known results for $\Sigma_{osc}$
for these cases \cite{curnoe2}, one easily verifies (\ref{eq:2dG})
and (\ref{eq:2dsigma}) again.  In fact the {\em scaling property}
(\ref{eq:2dsigma}), of $\Sigma_{osc}/\bar{\Sigma}$ as a function of 
$\omega_c/\mu$, depends only on the dimensionality of the
graph (as well as the presence of at least one internal fermion line
\cite{lutti2}), and is true of all higher graphs.

{\bf (ii) Alternative Expansions:}  There are two cases for 
which a simple alternative to  (\ref{eq:3dosc}) can be found
for $\Omega_{osc}$.

The first is where vertex corrections to the usual
Schwinger-Dyson/Nambu-Eliashberg self-energy (Fig.\ \ref{fi:feyn}(c)) can be 
neglected.  Then $\Sigma = \lambda^2 \int {\cal GD}$, where
${\cal G}$ and ${\cal D}$ are given self-consistently in terms
of $\Sigma$, thus summing over all ``rainbow graphs''.  The relevant
skeleton graph $\Phi_2$ (Fig.\ \ref{fi:feyn}(d)) then exactly cancels
$\beta^{-1} \mbox{Tr}[\Sigma\cal{G}]$ in (\ref{eq:luttiexp}), and
$\Omega = \bar{\Omega} +\Omega_{osc}$ is given,
{\em to all orders in} $\omega_c/\mu$, by
\begin{eqnarray}
\Omega &=& \frac{-1}{\beta} \mbox{Tr}[\log{\cal G}^{-1}] \\
  & = & \frac{-1}{\beta}\sum_{i\omega_m,n,\sigma}\log[
 i\omega_m - \epsilon_{\sigma n} -\Sigma(i\omega_m,\epsilon_{\sigma n})]
\label{eq:2dosc}
\end{eqnarray}
The crucial difference from (\ref{eq:3dosc}) (apart from the suppression
of $k_z$) is that $\Sigma$ now includes $\Sigma_{osc}$.  Deviations from
(\ref{eq:2dosc}) arise from ``crossed'' graphs (Fig.\ \ref{fi:feyn}(e)),
and there are many physical cases in which these are unimportant.
In the case of composite fermions the 
corrections from crossed ``gauge fluctuation'' graphs are not small, 
but at low energy their main effect both in zero field
\cite{gan1}, and finite field \cite{curnoe2}, is simply to renormalise
the vertices in the rainbow graph sum, without changing the functional form
of $\Sigma$. Thus
this approximation actually works well even beyond the ``Migdal limit'' in
which crossed graphs are small.
The difference between $\mbox{Tr}[\log\bar{\cal G}^{-1}]$ and
Tr$[\log{\cal G}^{-1}]$ depends crucially on how big is 
$\Sigma_{osc}/\bar{\Sigma}$; even though formally this is $\sim O(\omega_c/\mu)$
for all 2D systems, its actual value, for a given 
$\omega_c/\mu$, varies enormously between different systems.

The second case is of more academic interest; it arises when we may write
$\Omega$ in terms of a set of ``statistical quasiparticle'' (SQP)
energies \cite{balia1} $\varepsilon_{\sigma\nu}$ as
\begin{equation}
\Omega = \frac{-1}{\beta} \sum_{\nu,\sigma}
\log( 1 + e^{\beta(\mu - \varepsilon_{\sigma\nu})})
\label{eq:quasi}
\end{equation}
where the $\varepsilon_{\sigma\nu}$ are conventionally defined by a 
Landau expansion, are {\em real} \cite{note1}, and are {\em not} equal 
to the energies $E_{\sigma\nu}$ defind from the 1-particle
Green function by $E_{\sigma\nu} - 
\mbox{Re}\Sigma_{\sigma\nu}(E_{\sigma\nu}) = 0$.
The problem with (\ref{eq:quasi}) is that it relies on the usual
assumption that switching on interactions, in a system already
in Landau level states, does not reclassify the energy levels.
This is definitely not true for CF's, once the
gauge interactions are switched on \cite{note2}.

We now consider the new result (\ref{eq:2dosc}) in more detail.
Supressing the sum over spins, we
rewrite (\ref{eq:2dosc}) as an  integral,
with $\Sigma(x) \equiv \Sigma^{'}(x) + i \Sigma^{''}(x)$
\begin{equation}
\Omega = \frac{m \omega_{c}}{\Phi_0} \sum_{n=0}
     \int \frac{dx}{\pi} n_f(x)
  \tan ^{-1}(\phi(x,n))
\end{equation}
where $\phi(x,n) = \mbox{Im}G/\mbox{Re}G$.
 The prefactor $\frac{m \omega_{c}}{\Phi_0}$ is the 
Landau level degeneracy.  Defining 
$\epsilon = n \omega_{c}  - \mu$, the Poisson sum formula is used 
to separate
the oscillatory components of $\Omega$: 
\begin{eqnarray}
\Omega & = &  
 \frac{m\omega_c}{\Phi_0}
 \int \frac{dx}{\pi} n_f(x) \left[
\int_{-\mu}^{\infty}d\epsilon \tan^{-1}(\phi(x,\epsilon))
\right. \nonumber \\
 & & \left. \hspace{-.2in} -2 \sum_{k=1}^{\infty}\frac{(-1)^k}{2\pi k}
\int_{-\infty}^{\infty}d\epsilon  \mbox{Im}{\cal G}(\epsilon,x) 
 \sin
\left(\frac{2 \pi k(\epsilon+\mu)}{\omega_c}\right)  \right]
\label{eq:omega1}
\end{eqnarray}
where in the oscillatory (ie., $k > 0$) terms in the Poisson sum 
we have extended the limits of the $\epsilon$ integral to 
$\pm \infty$ and integrated
by parts.

\begin{figure}[ht]
\epsfysize=2.3in
\epsfbox[35 270 580 640]{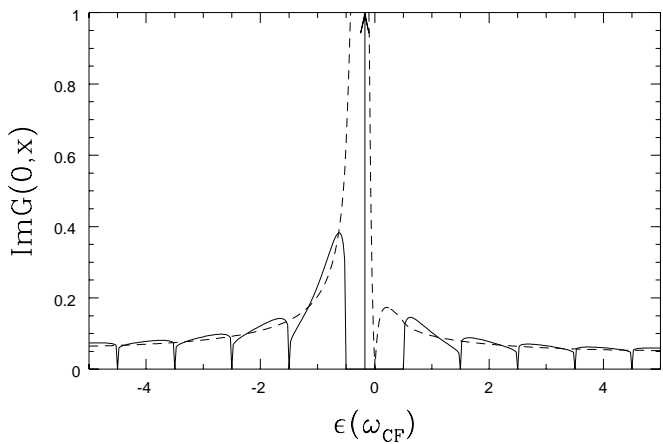}
\caption{Plot of the quasiparticle spectral function $\mbox{Im}{\cal G}(0,x)$
at zero temperature
for the case of composite fermions (with $\Sigma$
calculated to 2nd order in the gauge coupling). The solid line
shows the result using the full ${\cal G}$ (including all oscillatory terms),
whereas the dashed line uses only the non-oscillatory $\bar{\cal {G}}$. 
Note the
presence  of a gap, as well as an isolated pole, in the full
spectral function. These results are for $k_B T = 0$.
\label{fi:intspec}}
\end{figure}

As noted above, the important difference from previous expressions for 
$\Omega$ here is the inclusion of oscillatory terms in ${\cal G}$; 
in Fig.\ 2 we show the effect of this 
oscillatory structure in ${\cal G}$, for the particular case of 
CF excitations.

{\bf (iii) Magnetisation Oscillations:} The classic dHvA effect is in the
magnetisation oscillations; recently experiments have succeeded in seeing
these in {\it single layer} systems \cite{weigers}.
Taking the derivative of $\Omega_{osc}$, we get:
\begin{eqnarray}
M_{osc} & =&  \frac{-\partial \Omega}{\partial B} = M_1 + M_2 
\label{eq:magnetis} \\
M_1  & = & \frac{-2\pi\mu}{\Phi_0\omega_c} 
\sum_{k=1}^{\infty} (-1)^k \int \frac{dx}{\pi} n_f(x) \nonumber \\
& & \times \mbox{Re}\left[ \exp\left(\frac{2\pi ik}{\omega_c}
(x-\Sigma(x)+\mu)\right)\right] \label{eq:m11}\\
M_2  & = & \frac{-2\pi m}{\Phi_0}\sum_{k=1}^{\infty}(-1)^k
\int\frac{dx}{\pi}n_f(x) \nonumber \\
& & \times \mbox{Re}\left[\frac{\partial\Sigma(x)}{\partial B}
      \exp\left(\frac{2\pi ik}{\omega_c}(x-\Sigma(x)+\mu)\right)\right] 
    \nonumber \\
& & -\frac{m}{\Phi_0}\int_{-\mu}^{\infty} d\epsilon \int \frac{dx}{\pi}n_f(x)
    \mbox{Im}G(\epsilon,x)\frac{\partial \Sigma}{\partial B} 
\label{eq:m21}
\end{eqnarray}
At first glance the first term  $M_1$ resembles the results of 
Fowler and Prange \cite{fowle1},
and Engelsberg and Simpson \cite{engel2}; however it now involves the {\it 
full} $\Sigma$ (including $\Sigma_{osc}$).
The second term $M_2$  is formally of the same order in $\omega_c/\mu$ 
as $M_1$, and quite new. Typically the term in 
$\frac{\partial \Sigma}{\partial B}$
dominates $M_2$, and we shall see below that in 2 dimensions it can be much
larger than $M_1$.

Eqs. (\ref{eq:magnetis}-\ref{eq:m21}) are
 valid for any 2-dimensional charged system for which
the full self-energy (including oscillatory contributions) can be written
down. 

{\bf (iv) Application to the Composite Fermion System:}
We now wish to demonstrate on a particular example that the deviations from
orthodox behaviour can be rather large. We choose the CF gauge theory, for
which 
the oscillatory self-energy for composite fermions interacting 
with gauge fluctuations was previously calculated \cite{curnoe1,curnoe2}. Here 
we assume unscreened Coulomb interactions (ie., the 
dynamical exponent \cite{hlr,stern1,kim1,curnoe1,curnoe2} is $s = 2$). 
For numerical work it is convenient to use a Matsubara sum over 
$\Sigma(x)$ evaluated at $x = i\omega_l = i\pi(2l+1)/\beta$; writing 
$\mbox{Im}\Sigma(i\omega_l) \equiv \xi(i\omega_l)$, we have 
\begin{eqnarray}
\xi(i\omega_l)& =&  2 \kappa_2 \left[
\frac{\omega_l}{\pi}\log\left(\frac{\omega_l}{\mu}\right)
+\frac{4}{\beta}\sum_{k^{'}=1}^{\infty}(-1)^{k^{'}}\sum_{\omega_m>0}
  \right. \nonumber \\
&& \hspace{-.4in}   \left. 
\exp\left(\frac{-2\pi k^{'} \omega_m}{\omega_{CF}}\right)\log\left(
\frac{\omega_l+\omega_m}
{\omega_{CF}}\right)
\cos\left(\frac{2\pi k^{'} \mu}{\omega_{CF}}\right)\right]
\label{eq:selfe5}
\end{eqnarray}
%\begin{eqnarray}
%\Sigma(x) = \frac{2\kappa_2}{\pi}\left[
%   x\log\left(\frac{x}{\mu}\right) -\frac{i|x|\pi}{2}
%+\sum_{k^{'}=1}^{\infty}\int_{-\mu}^{\infty} 
%\cos\left(\frac{2\pi k^{'} (\mu+\epsilon)}\right)
%\left(\log|\epsilon-x|
%  (n_f(-\epsilon) - n_f(\epsilon))
%{\omega_{CF}}\right) +i 
where the coupling $\kappa_2$ is usually slightly less than one \cite{curnoe2}.
In QO experiments one 
examines $\log(A)$; 
in LK theory $\log(A)$ should be a linear function of
$1/B$ (the ``Dingle plot''), as well as of $T$ (the ``mass plot''). 
Fig.\ 3 shows (for the example of CF fermions), the importance of  
$M_2$, as well as the 
considerable non-linearity shown in Dingle plots (which we also find in the
mass plots, not shown here). 
Thus in this example a conventional analysis of QO phenomena, using either  
the LK formula, or its generalisations \cite{fowle1,engel2} clearly fails. 
We do not believe this example to be untypical (in fact if we choose 
{\it screened} short-range interactions between the 
CF's \cite{hlr,curnoe1,curnoe2}, 
with dynamical exponent $s = 3$, we get much worse deviations!).
We thus believe 
that where strong violations of conventional behaviour are
observed \cite{Balthes}, or where interaction effects are known to be strong
\cite{weigers}, one should re-analyse the data using the results herein. In
the context of the FQHE near half-filling, fits of 
QO results to LK theory should clearly be treated with caution.

\begin{figure}[ht]
\epsfysize=2.7in
\epsfbox[35 250 580 690]{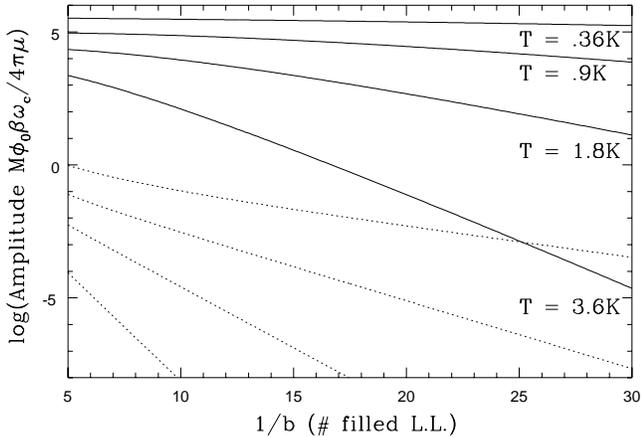}
\caption{A numerical evaluation of $\log(A)$, where $A$ is the 
amplitude of the dHvA oscillations in $M$, as a function of
$1/B$ for various fixed temperatures, for a system of CF quasiparticles
(we assume an unscreened Coulomb interaction, and use the 
second-order result for $\Sigma$ derived earlier [14].
We assume a chemical
potential of 180K, and a coupling constant $\kappa_2 = 0.8$.  
The dashed lines show 
$\log(A_1)$, 
 and the solid lines show $\log(A_1 + A_2)$, where $A_1$ and $A_2$ are
 the amplitudes of oscillations of $M_1$ and $M_2$. 
\label{fi:Mag}}
\end{figure}

In summary, we have shown that the LK formalism (or its 
many-body generalisations \cite{fowle1,engel2})
for describing quantum oscillations {\em breaks down} in two dimensions.
To remedy this, we have derived new results that 
can be applied when 
crossed diagrams may be neglected. We have applied these results  
to a problem of current interest \cite{willett}, ie.,
composite fermions interacting with gauge 
fluctuations (believed to give a good description of the fractional 
quantum Hall states, at least near half-filling). The results show 
radical departures from LK behaviour. Such departures should also 
exist in other strongly-interacting 2-dimensional electronic 
systems, whether or not
they behave as Fermi liquids in zero field.

P. C. E. S. thanks G. Martinez, I. D. Vagner, and P. Wyder, for hospitality 
and support in 
Grenoble, as well as 
the CIAR and NSERC of Canada.  S. C. acknowleges support
from the Feinberg Graduate School of the Weizmann Institute.

\vspace{.2in}
$^*$present address


\begin{references}
\bibitem{gauge} It is sometimes objected that the self-energy, not 
being gauge invariant, should not be used in formulae for physical 
quantities like $\Omega$ or $M$ (which are). We simply note here that the
gauge dependence of $\Sigma$ is irrelevant in the formulas in this paper
since one {\it integrates over} $\Sigma$; moreover typically only the 
pole structure in the Green function is relevant, and this is also gauge
invariant (cf. Stern and Halperin \cite{stern1}).
\bibitem{note1}  If we put Im$\bar{\Sigma} = 0$ in Eq. (\ref{eq:3dosc}),
   then we get Eq. (\ref{eq:quasi}) with $E_{\sigma\nu}$ instead of 
   the SQP energy $\varepsilon_{\sigma\nu}$.  However, as shown in 
   [14] and [15], even in conventional theory one
  should not drop Im$\bar{\Sigma}$.  The difference between 
  $\varepsilon_{\sigma\nu}$ and $E_{\sigma\nu}$ is discussed in great
  detail by Pethick and Carneiro \cite{balia1} in the zero-field case where 
  the quantum number is $r \equiv k$, the momentum.
\end{references}
\end{document}